\documentclass[prb,
12pt,
superscriptaddress,showpacs,amsmath,amssymb]{revtex4}
\usepackage{amsfonts}
\usepackage{bm}
\usepackage{verbatim}

\usepackage{graphicx}

\begin{document}

\author{G.E.~Volovik}
\affiliation{Landau Institute for Theoretical Physics, acad. Semyonov av., 1a, 142432,
Chernogolovka, Russia}

\title{Thermodynamics of Einstein static Universe with boundary}

\date{\today}

\begin{abstract}
The de Sitter state and the static Einstein Universe are unique states that have a constant scalar Ricci curvature ${\cal R}$. It was shown earlier that such a unique symmetry of the de Sitter state leads to special thermodynamic properties of this state, which are determined by the local temperature $T=1/(\pi R)$, where $R$ is the radius of the cosmological horizon.
Then, what happens in the static Universe? We consider the original Einstein Universe, i.e. the so-called spherical Universe, which is the half of the elliptical $S^3$ Universe and has a boundary at $r=R$. This boundary is viewed as a physical surface that connects the Einstein Universe with the thermal environment. In this realization, the Einstein Universe, it is characterized by a local temperature $T=1/(\pi R)$, which is analogous to the local temperature of the de Sitter state. Or, conversely, the temperature of environment heat bath determines the radius of the Universe, $R=1/\pi T$. The thermodynamics of the bounded Einstein Universe is also analogous to the thermodynamics of the de Sitter state. In particular, the entropy of the elliptical Universe $R\times S^3$, which splits into two Einstein Universes, satisfies the holographic relation, $S=A/4G$, where $A=2\pi^2R^2$ is the area of the boundary. This shows that in thermodynamics this boundary plays the same role as the cosmological de Sitter horizon. In this Universe, Zeldovich stiff matter is also preferred.
\end{abstract}
%\pacs{
%}

\maketitle

\newpage
\tableofcontents

\section{Introduction}
\label{IntroductionSec}

Recently the possible connection between the entropy of the de Sitter state and the entropy of the static Einstein Universe has been discussed and it was found that these entropies can be parametrically equal.\cite{Kaloper2024} Here we consider the exact connection, in which the entropy of both states has the Bekenstein-Hawking value $A/4G$. Here $A$ is the area of the cosmological horizon in the de Sitter case. In the $S^3$ Universe, $A$ is the area of the equator, which serves as the physical boundary separating two static Einstein Universes. This boundary plays the same role as the de Sitter cosmological horizon. The connection between the thermodynamics of the de Sitter state and the static Universe follows from the specific symmetries of both states, due to which the curvature is constant in space and time. The constant curvature provides the constant local temperature in both systems.

In the de Sitter state this is the symmetry of the space-time with respect to the combined translations, which in the Minkowski limit becomes the invariance of under translations. This symmetry gives rise to the local thermodynamics of the de Sitter state, which is characterized by the local temperature.\cite{Volovik2024f,Volovik2024g} 
This consideration is based on observation, that matter immersed in the de Sitter vacuum feels the de Sitter vacuum as the heat bath with the local temperature $T=H/\pi$, where $H$ is the Hubble parameter, see also Ref.\cite{Maxfield2022}. This temperature is twice the Gibbons-Hawking temperature, $T_{\rm GH}=H/2\pi$. But this leads to the local thermodynamics with the entropy density, which gives rise to the holographic connection between the bulk entropy of the Hubble volume and the surface Gibbons-Hawking entropy  of the cosmological horizon, $S_{\rm bulk}=S_{\rm surface}= A/4G$.

In the static Einstein $R\times S^3$ Universe, the curvature is also constant in space and time.
That is why one may expect that it may have the similar properties. But what is the analog of the cosmological event horizon? 

The static Universe is unstable,\cite{Eddington1930} but it can be stabilized if it is connected with the external environment -- the analog of the external heat bath.\cite{Barcelo2004} We consider one of the possible channel of the connection of the static Universe with the environment. In this example, the $S^3$ Universe splits into two half spheres, each having boundary with the environment. Each part is the original   static Universe considered by Einstein,\cite{Einstein1917}  see also the review \cite{Einstein2017}. 

 For such splitted Universe with boundary we obtain that the radiation rate corresponds to the radiation in the Minkowski state in the presence of the heat bath with temperature  $T=1/\pi R$, where $R$ is the radius of the static Universe. This suggests that on one hand the temperature  $T=1/\pi R$ represents the local temperature of the bounded Einstein Universe, which determines the thermodynamic of the Universe. On the other hand the temperature of the environment determines the radius of the static Universe, $R=1/\pi T$. This gives the holographic connection $S_M=A/4G$ between the entropy $S_M$ of matter and the area of the boundary, $A=2\pi^2R^2$. This demonstrates that this boundary plays the same role as the de Sitter cosmological horizon. Thermal contact between the static Universe and the surrounding heat bath favours Zeldovich stiff matter, which also has relation to the de Sitter thermodynamics.

\section{Static Universe vs de Sitter}
\label{StaticDS}

\subsection{Equilibrium conditions}
\label{EquilibriumSec}

Both the de Sitter state and the Einstein static Universe have constant in space scalar curvature ${\cal R}$. The curvature serves as the thermodynamic variable, which is responsible for the gravitational degrees of freedom. It can be considered in the same way as the thermodynamic variables describing the matter degrees of freedom. One may introduce the stress-energy tensor for gravitational field as the variation of the gravitational term (the curvature term) in the Einstein action.\cite{Volovik2003} For the finite in space system, the variation of the total Einstein action can be written in terms of stress-energy tensors  $T_{\mu\nu}^M$, $T_{\mu\nu}^\Lambda$ and $T_{\mu\nu}^G$ of the three components: matter, vacuum energy and gravity: 
\begin{eqnarray}
 \delta S= \frac{1}{2} \int d^4x \sqrt{-g}\left( T_{\mu\nu}^M +T_{\mu\nu}^\Lambda+T_{\mu\nu}^G\right) \delta g^{\mu\nu}\,.
\label{ActionVariation}
\end{eqnarray}
The standard Einstein equation, $\delta S=0$, demonstrates that in this approach the total stress-energy tensor is zero:
\begin{eqnarray}
T_{\mu\nu}=T_{\mu\nu}^M +T_{\mu\nu}^\Lambda+T_{\mu\nu}^G=0\,,
\label{TotalEnergyMomentum}
\end{eqnarray}
and thus the conservation of energy and momentum is automatically fulfilled: 
$T^{\mu\nu}_{,\nu}=T^{\mu\nu}_{;\nu}=0$.

For static Einstein Universe, the gravitational part $T_{\mu\nu}^G$ of the stress-energy tensor
is expressed via the Ricci scalar curvature ${\cal R}$. Then, if the curvature is the space-time constant, the Eq.(\ref{TotalEnergyMomentum}) can be written in terms of the energy densities and partial pressures of the three components (see Eq.(29.38) in Ref. \cite{Volovik2003}):
\begin{eqnarray}
\rho=\rho_M +\rho_\Lambda +  \rho_R=0\,,
\label{EinsteinUn1}
\\
P=P_M +P_\Lambda +  P_R=0\,.
\label{EinsteinUn2}
\end{eqnarray}
Here $P_M=w_M \rho_M$ is the equation of state of matter (we assume for simplicity that there is only one type of matter components, such as cold matter with $w_M=0$ or radiation with $w_M=1/3$); $P_\Lambda=-\rho_\Lambda$ is the equation of state of the dark energy. The quantity $\rho_R$ is the energy density stored in the scalar curvature, and $P_R=w_R \rho_R$ is the partial pressure related to the space curvature with $w_R=-1/3$ in the Einstein Universe, see below.

The physical meaning of equation (\ref{EinsteinUn1}) is the gravineutrality of the system, which is the analogue of electroneutrality. The equation (\ref{EinsteinUn2}) means the absence of external pressure acting on the Universe. In this case the partial pressures of the three components must cancel each other.

\subsection{Equilibrium conditions for de Sitter state}
\label{EquilibriumDSSec}

For the de Sitter state, the equations (\ref{EinsteinUn1}) and (\ref{EinsteinUn2}) give $w_R=w_\Lambda=-1$ and $\rho_\Lambda= -\rho_R$. But this is wrong, since the de Sitter state is infinite with the cosmological horizon, and thus the above considerations are not applicable. Instead, the gravitational energy density and the corresponding partial pressure are:
\begin{eqnarray}
\rho_R= \frac{1}{2}K{\cal R}=6KH^2 \,,
\label{CurvatureEnergyDS}
\\
P_R =- \frac{d (\rho_R V)}{dV} +K {\cal R} =  \rho_R \,.
\label{CurvaturePressureDS}
\end{eqnarray}
Here $H$ is the Hubble parameter and $K$ is the gravitational coupling:
\begin{equation}
 K= \frac{1}{16\pi G}
 \,.
\label{KG}
\end{equation}
In Eq. (\ref{CurvaturePressureDS}) we take into account that the energy of the vacuum in the arbitrary volume $V$ is proportional to $V$. This equation demonstrates that the equation of state of the gravitational component in the de Sitter state is the same as equation of state of Zel'dovich stiff matter:
\begin{equation}
w_R=+1 \,\,,\,\, w_\Lambda=-1
 \,.
\label{2components}
\end{equation}

Since de Sitter state does not contain matter, the equilibrium conditions for de Sitter is
\begin{equation}
 P=P_\Lambda +  P_R=0 
 \,.
\label{dSConditions}
\end{equation}
This condition means the absence of the external pressure and it gives $\rho_\Lambda= 6KH^2$.

\subsection{Equilibrium conditions for static Universe}
\label{EquilibriumStaticSec}

As distinct from de Sitter, the static Universe is finite, and the equations (\ref{EinsteinUn1}) and (\ref{EinsteinUn2}) are applicable. The equation of state for gravity with space curvature is $w_R=-1/3$:
\begin{eqnarray}
\rho_R= -K{\cal R}=-\frac{6K}{R^2} \,,
\label{CurvatureEnergyStatic}
\\
P_R =- \frac{d (\rho_R V)}{dV}=- \frac{1}{3} \rho_R= \frac{2K}{R^2} \,.
\label{CurvaturePressureStatic}
\end{eqnarray}
Here we take into account that the volume of the static Universe is proportional to $R^3$, where $R$ is the radius of the Universe, while $\rho_R \propto 1/R^2$ .

Then the equilibrium conditions  (\ref{EinsteinUn1}) and (\ref{EinsteinUn2}) give the following matter density and vacuum energy density in the static Universe:
\begin{eqnarray}
\rho_M = \frac{2}{3(1+w_M)}  K{\cal R}\,,
\label{Matter}
\\
\rho_M + p_M= \frac{2}{3}  K{\cal R}=\frac{4K}{R^2}\,,
\label{MatterP}
\\
\rho_\Lambda=  K{\cal R} \left( 1- \frac{2}{3(1+w_M)}  \right)\,.
\label{Vacuum}
\end{eqnarray}

\subsection{Particular states of static Universe}
\label{States}

It was shown in Ref. \cite{Volovik2024f}, that the thermodynamics of the de Sitter state has some properties of Zel'dovich stiff matter with $w_M=1$. For the static Universe with $w_M=1$, one has the following relation between the vacuum energy density and the radius $R$ of this Universe:
\begin{equation}
\rho_\Lambda(w_M=1)=2\rho_M =\frac{2}{3}K{\cal R}=\frac{1}{4\pi GR^2}
 \,.
\label{Zeldovich}
\end{equation}
In case of the cold matter with $w_M=0$:
\begin{equation}
\rho_\Lambda(w_M=0)=\frac{1}{2}\rho_M =\frac{1}{3}K{\cal R}=\frac{1}{8\pi GR^2}
 \,.
\label{cold}
\end{equation}

The special case is $w_M=-1/3$, at which the equation of state for matter is the same as the equation of state for curvature, $w_M=w_R=-1/3$. In the expanding Universe this equation of states is on the border between deceleration and acceleration, see e.g. Ref. \cite{Odintsov2024}. In the static Einstein Universe this corresponds to the absence of the dark energy component:\cite{Maeda2024,Whittaker1968}
\begin{equation}
\rho_\Lambda(w_M=w_R=-1/3)=0
 \,.
\label{special}
\end{equation}
This is similar to the 2-component situation in de Sitter in Eq.(\ref{2components}), where the matter component is absent.

\section{Creation of massive particles and local temperature}
\label{CreationSec}

\subsection{Creation of massive particles in the Universe with boundary}

In the spherical Universe, the metric is given by:
\begin{equation}
ds^2=-dt^2 + \frac{dr^2}{1 - \frac{r^2}{R^2}} + r^2 d\Omega^2 \,,
\label{Metric}
\end{equation}
which can be compared with the de Sitter metric with $R=1/H$:
\begin{equation}
ds^2=-dt^2\left(1 - \frac{r^2}{R^2}\right) + \frac{dr^2}{1 - \frac{r^2}{R^2}} + r^2 d\Omega^2 \,.
\label{dSMetric}
\end{equation}

Both metrics have coordinate singularity at $r=R$. This singularity can be either unphysical, which can be removed by the proper coordinate transformation, or physical if it serves as the boundary of the Universe. 
 
The round metric on the 3-sphere has no singularities, and it explicitly demonstrates the uniform nature of the static Universe:
\begin{equation}
dl^2=d\eta^2 + \sin^2\eta \,d\xi_1^2+ \cos^2\eta \, d\xi_2^2\,,
\label{metric}
\end{equation}
where $0<\eta <\pi/2$ and $0 <\xi_1,\xi_2 <2\pi$.

The volume $V$ of the $S^3$ Universe is:
\begin{equation}
V=2\pi^2 R^3\,.
\label{Volume}
\end{equation} 
This Universe is also characterized by the 2D "equator" at $\eta=\pi/4$, which area $A$ is:
\begin{equation}
A=2\pi^2 R^2\,.
\label{Area}
\end{equation}
Let us recall that the intersection of a 3D-sphere with a three-dimensional hyperplane is a 2D-sphere. As the 3D-sphere moves
through the 3D-hyperplane, the intersection represents the growing 2D-sphere that reaches its maximal size  and then shrinks. The maximum size corresponds to the "equator" with the area $A$ in Eq.(\ref{Area}).

The physical singularity takes place in case of the spherical Einstein Universe with the boundary at the equator. The volume of this Universe is half of the total volume $V$ of the $S^3$: 
\begin{equation}
V_\frac{1}{2} = 4\pi \int_0^R dr \frac{r^2}{\sqrt{1 - \frac{r^2}{R^2}}} =\pi^2 R^3 =\frac{V}{2}\,.
\label{HalfVolume}
\end{equation}

 Let us consider the quantum tunneling processes of the creation of massive particles in the two Universes: the Einstein Universe with boundary and the de Sitter Universe with the cosmological horizon.
 In the semiclassical limit the tunneling trajectory $p_r(r)$ of particle with mass $M$ in the Einstein Universe is determined by equation:
 \begin{equation}
E^2=g^{rr}p_r^2 +M^2\,\,,\,\, g^{rr}=1 - \frac{r^2}{R^2}
 \,,
\label{TunnelingTrajectory}
\end{equation}
while in the static de Sitter one has
 \begin{equation}
 \frac{E^2}{1 - \frac{r^2}{R^2}} =g^{rr}p_r^2 +M^2\,\,,\,\, g^{rr}=1 - \frac{r^2}{R^2}
 \,.
\label{dSTrajectory}
\end{equation}

For zero energy of the created particle, $E=0$, one obtains the same trajectory in both Universes:
\begin{equation}
p_r(r) = i \frac{M}{\sqrt{1 - \frac{r^2}{R^2}}}
 \,.
\label{CommonTrajectory}
\end{equation}
Since the rate of particle creation is given by the imaginary part of the action along this trajectory, then  in the WKB approximation one would obtain the following creation rate:
\begin{equation}
w \sim  \exp(-2 {\rm Im} \,S)  = \exp\left( - 2M\int_0^R \frac{dr}{\sqrt{1 - \frac{r^2}{R^2}}} \right) = \exp(-\pi MR)
 \,.
\label{CreationRate1}
\end{equation}
 The rate is the same as the rate of particle creation in the thermal bath with temperature $T=1/(\pi R)$ in the flat space:
 \begin{equation}
w \sim \exp\left(-\frac{M}{T}\right) \,\,,\,\, T=\frac{1}{\pi R}
 \,.
\label{CreationRate}
\end{equation}
For the de Sitter state, this have been obtained using the Painleve-Gullstrand metric, which has no coordinate singularity at the horizon. This is also the natural result in the static de Sitter coordinates, since the trajectory can be extended to the region behind the horizon, where the momentum $p_r(r)$ becomes real.

This quantum tunneling process is impossible for the static elliptical Universe with the $S^3$ space, since the region with $r>R$ does not exist. But this process becomes possible if the static Universe represents 
only the half of the $R\times S^3$ Universe. It can be for example the part of the larger Universe, where in the region at $r>R$ the metric has the opposite signature. 
In this case we obtain the same Eq.(\ref{CreationRate1}) for the creation rate, and thus the same temperature $T=1/\pi R$. This temperature is fully determined by the radius $R$ of the Universe regardless of the equation of state $w_M$ for matter. 

In this approach, the boundary at $r=R$ in this half-Universe plays the same tole as the cosmological horizon of the de Sitter state. Also, in the same way as the creation of particles by the de Sitter heat bath leads to the instability and decay of the de Sitter Universe,\cite{Volovik2024f,Volovik2024g} the creation of particles in the Einstein Universe with boundary leads its decay. 

\section{Thermodynamics of static Universe and holographic principle}
\label{ThermodynamicsSec}

However, the equilibrium state is possible, if the environment serves as the heat bath. In this case the temperature of the heat bath determines the radius of the static Universe, $R=1/\pi T$, and its thermodynamics.
 Using Eq.(\ref{MatterP}) and the Universe radius $R=1/\pi T$,
one obtains 
\begin{equation}
\rho_M +P_M= 4\pi^2KT^2\,,
\label{GibbsDuhem}
\end{equation}

According to the Gibbs-Duhem relation the right-hand side  is equal to $Ts_M$, where $s_M$ is the entropy density of matter. One obtains
\begin{equation}
s_M =4\pi^2KT=\frac{4\pi K}{R}=\frac{1}{4GR}
 \,.
\label{EntropyDensity}
\end{equation}
Then the total entropy of the bounded Universe in the volume $V_\frac{1}{2}=V/2=\pi^2R^3$: 
\begin{equation}
S=s_M V_\frac{1}{2}=\frac{1}{4GR} \,\pi^2 R^3=\frac{\pi^2 R^2}{4G} =\frac{A}{8G}\,.
\label{holography1}
\end{equation}

This entropy looks as the analog of holographic area law similar to that for the entropy of the Hubble volume in de Sitter Universe. This actually supports the possible connection between Bekenstein-Hawking entropy of the 2+1 de Sitter horizon and that of static Einstein $R\times S^2$ Universe.\cite{Kaloper2024}
But now the role of the cosmological horizon is played by the boundary of the static half-Universe. 

To get the full correspondence with the holographic principle one should consider the entropy of both parts of Einstein Universe with the total volume $V$:
\begin{equation}
S=s_M V=\frac{1}{4GR} \,2\pi^2 R^3=\frac{\pi^2 R^2}{2G} =\frac{A}{4G}\,.
\label{holography}
\end{equation}
The role of the horizon is played by the 2D equator of the $S^3$ space, which separates two half-Universes and serves as the boundary which connects these Universes with environment. 

The holographic relation in Eq.(\ref{holography}) dictates the equation of state for matter. The linear in $T$ temperature dependence of the entropy density of matter takes place only for matter with $w_M=1$, i.e. for the Zeldovich stiff matter. This would mean that the equilibrium between the static Universe and its environment is only possible if all matter is stiff. For matter with $w_M \neq 1$ the exchange with environment leads to the decay of the static Universe.

\section{Conclusion}
\label{ConclusionSec}

The local temperature, $T=1/\pi R$, and holographic connection for the entropy,  $S=A/4G$, make sense in the static Einstein Universe if the Universe is connected with the external environment and can radiate particles. We considered the rate of radiation using as an example the spatial metric in Eq.(\ref{Metric}), which contains the coordinate singularity. This corresponds to an $S^3$ Universe divided into two half-universes, where the coordinate singularity represents a real physical boundary. The rate of radiation of particles is the same as the rate of radiation in Minkowski state in the presence of the heat bath with temperature $T=1/\pi R$.

The equilibrium between the heat bath and the Einstein Universe is possible if the environment has a finite temperature. In this case the temperature of environment determines the radius of the Universe, $R=1/\pi T$. However, the full equilibration is only possible for the Universe with Zeldovich stiff matter with equation of state $w=1$.

The same temperature and the same holographic law take place for the de Sitter state, where the role of the boundary is played by the cosmological horizon with $R=1/H$. The reason is that both states can be described by the same spatial part of metric. In both states the curvature is constant in space and time, and serves as the thermodynamic variable together with the inverse Newton "constant", which is thermodynamically conjugate to curvature. Moreover, in both states the important role is played by the Zeldovich stiff matter with the same temperature $T=1/\pi R$.

This observation confirms the connection between the entropy of the Bekenstein-Hawking horizon in the 2+1 de Sitter and the entropy of the static Einstein $R\times S^2$ universe.\cite{Kaloper2024} But instead of parametric agreement, we have an exact match, which suggests that our approach may make sense.

The open question concerns the generalized statistics, which usually follows the holographic connection.
For example, the black holes in the Minkowski environment are the finite systems which entropies obey the composition law described by the non-extensive Tsallis-Cirto statistics.
\cite{Volovik2025,Volovik2025a,Odintsov2025}  The Tsallis-Cirto $\delta=2$ statistics correctly determines the entropy deficit in the processes of splitting and merging of black holes,
$S(M_1 +M_2) - S(M_1) -S(M_2) =2 \sqrt{S(M_1)S(M_2)}$. The entropy here is determined not by the number of degrees of freedom, but by the number of the elements of the matrix of the correlations between these degrees of freedom. This reminds the Matrix Theory suggested by Susskind.\cite{Susskind2023}

On the other hand, the de Sitter entropy of the unbounded de Sitter state is extensive, being proportional to the volume of the considered part of the Universe, so that $S(V_1+V_2)=S(V_1) + S(V_2)$. 
The situation when the black holes are in the de Sitter environment with its own cosmological horizon is more complicated. The composition law for the entropy deficit in the processes which combine  the black hole horizons and the cosmological horizon remains unclear. There are some hints\cite{Susskind2023} that the Matrix Theory can be valid also in the de Sitter environment. Can the Matrix Theory be extended to the entropy of Einstein static Universe with boundaries also remains an open question.

\end{document}